\documentclass{caosp306}

\usepackage{graphicx}

\usepackage{natbib}
\articleNo{123}
\pubyear{2005}
\volume{35}
\volnumber{3}
\firstpage{1}
\received{May 1, 2005}
\accepted{August 28, 2005}

\def\BibTeX{{\rm B\kern-.05em{\sc i\kern-.025em b}\kern-.08em
             T\kern-.1667em\lower.7ex\hbox{E}\kern-.125emX}}

\begin{document}

%
\hauthor{L.\,Szabados}

\title{Selected new results on pulsating variable stars}


%
\author{
        L.\,Szabados 
       }

%
\institute{
Konkoly Observatory, Research Centre for Astronomy and Earth Sciences of the
HAS, MTA CSFK Lend\"ulet Near-Field Cosmology Research Group, 
H-1121 Budapest, Konkoly Thege Mikl\'os \'ut 15-17, Hungary
          }
\date{October 30, 2018}

\maketitle

\begin{abstract}
Recent progress in the studies of pulsating variable stars is summarized
from an observational point of view. A number of unexpected phenomena have been 
revealed in the case of pulsators in the classical instability strip. 
These discoveries -- lacking theoretical explanation yet -- make pulsating stars
more valuable objects for astrophysics than before. The emphasis is laid on 
Cepheids of all kind and RR~Lyrae type variables, as well as binarity among
pulsating variable stars.
\keywords{pulsating variables -- radial pulsation -- nonradial pulsation --
binarity}
\end{abstract}

%
\section{Introduction}
\label{intr}

This paper is intended to be an update and continuation of the
review published in the proceedings of the previous conference 
held in Tatransk\'a Lomnica on similar topic five years ago 
(Szabados, 2014). The reader will find repetitions only in unavoidable
cases. 

\begin{figure}[!]
\centerline{\includegraphics[width=6.0cm,clip=]{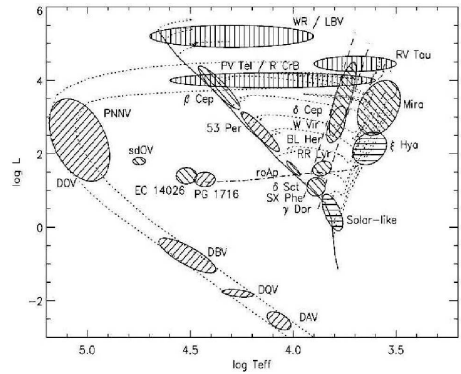}}
\caption{H-R diagram showing the location of various types of pulsating 
variables (Jeffery, 2008). Some more types are listed in Table~1.}
\label{fig1}
\end{figure}

\begin{table}  
\footnotesize
\begin{center}  
\caption{Classification of pulsating variable stars.} 
\label{pulsvartypes}  
\begin{tabular}{l@{\hskip2mm}l@{\hskip2mm}l@{\hskip2mm}r@{\hskip2mm}c@{\hskip2mm}l} 
\hline\hline\\
Type &  Design. & Spectrum & Period & Amplitude & Remark$^\ast$\\
 & & & &in $V$ band (m) & \\[0.2ex]
\hline
Cepheids    & DCEP   &  F-G Iab-II  & 1-135\,d   &  0.03-2  & \\
            & DCEPS  &  F5-F8 Iab-II &  $<$7\,d  &  $<$0.5  & 1OT \\
	    & CEP(B) &  F5-F6 Iab-II &  2-7\,d   &  0.1-1   & beat Cepheids \\
BL Boo      & ACEP   &  A-F         & 0.4-2\,d   &  0.4-1.0 & anomalous Cepheids\\
W Vir       & CWA    &  FIb         &    $>$8\,d &  0.3-1.2 & \\
BL Her      & CWB    &  FII         &    $<$8\,d &  $<$1.2  & \\
RV Tau      & RV, RVA&  F-G         & 30-150\,d  &  up to 3 & \\
            & RVB    &  F-G         & 30-150\,d  &  up to 3 & var. mean brightness\\
RR Lyr      & RRAB   &  A-F giant   & 0.3-1.2\,d &  0.4-2   & \\
            & RRC    &  A-F giant   & 0.2-0.5\,d & $<$0.8   & 1OT\\
	    & RR(B)  &  A-F giant   & 0.2-1.0\,d & 0.4-2    & double-mode puls.\\
$\delta$ Sct & DSCT  &  A0-F5\,III-V & 0.01-0.2\,d & 0.003-0.9 & R+NR\\
SX Phe      & SXPHE  &  A2-F5 subdw.& 0.04-0.08\,d &$<$0.7  &  Pop. II\\
$\gamma$ Dor & GDOR  &  A7-F7\,IV-V & 0.3-3\,d  & $<$0.1   &  NR, low degree g-mode \\
roAp        & ROAP   &  B8-F0\,Vp   & 5-20 min   & 0.01    &  NR p-modes\\
$\lambda$ Boo & LBOO &  A-F         & $<$0.1\,d  & $<$0.05 &  Pop.\,I, metal-poor\\
Maia        &        &  A           &            &         &  to be confirmed\\
V361 Hya    & RPHS,  &  sdB         & 80-600\,s  & 0.02-0.05 & NR, p-mode\\
            & EC14026 &             &            &         & \\
V1093 Her   & PG1716, & sdB         & 45-180 min & $<$0.02 &   g-mode\\
            & Betsy  &	            &            &         & \\
DW Lyn      &        &  subdwarf    &            & $<$0.05 &   V1093\,Her\,+\,V361\,Hya\\
GW Vir      & DOV,   &  HeII, CIV   & 300-5000\,s & $<$0.2 &   NR g-modes \\
            & PG1159 &              &            &         & \\   
ZZ Cet      & DAV    &  DAV         & 30-1500\,s & 0.001-0.2 & NR g-modes\\
DQV         & DQV    &  white dwarf & 7-18 min   & $<$0.05 & hot carbon atmosphere\\
V777 Her    & DBV    &  He lines    & 100-1000\,s &$<$0.2  & NR g-modes\\[0.3ex]
\hline
Solar-like  &        & F5-K1\,III-V & $<$hours  & $<$0.05  & many modes\\
\hspace*{5mm}oscill. &     &        &           &          & \\[0.3ex]
Mira        & M      &  M, C, S IIIe & 80-1000\,d & 2.5-11 & small bolometric ampl.\\
Small ampl. & SARV   &  K-M\,IIIe   & 10-2000\,d  & $<$1.0 & \\
\hspace*{5mm}red var.  &        &              &            &         & \\
Semi-regular & SR    &  late type I-III & 20-2300\,d & 0.04-2 & \\
            &  SRA   & M, C, S\,III & 35-1200\,d & $<$2.5  & R overtone\\
            &  SRB   & M, C, S\,III & 20-2300\,d & $<$2    & weak periodicity\\
	    &  SRC   & M, C, S\,I-II & 30-2000\,d & 1      & \\
            &  SRD   & F-K\,I-III   & 30-1100\,d & 0.1-4   & \\	    
            &  SRS   & late type    & 5-30\,d & 0.1-0.6    & high-overtone puls.\\
Long-period &  L     & late type    &            &         & slow \\
\hspace*{5mm}irregular &   &        &            &         & \\
            &  LB    & K-M, C, S III &           &         & \\
            &  LC    & K-M I-III    &            &         & \\
Protoplan.  & PPN    &  F-G I       & 35-200\,d   &        & SG, IR excess\\
\hspace*{5mm}nebulae &  &           &             &        & \\[0.3ex]
\hline
\end{tabular} 
\end{center}  
$^\ast$ R = radial; NR = non-radial; 1OT = first overtone; SG = supergiant.
Spectrum is given for maximum brightness for large amplitude variables.
\end{table}

\setcounter{table}{0}
\begin{table}  
\footnotesize
\begin{center}  
\caption{Classification of pulsating variable stars (continued).} 
\begin{tabular}{l@{\hskip2mm}l@{\hskip2mm}l@{\hskip2mm}r@{\hskip2mm}c@{\hskip2mm}l} 
\hline\\
Type &  Design. & Spectrum & Period & Amplitude & Remark$^\ast$\\
 & & & &in $V$ band (m) & \\[0.2ex]
\hline
53 Per      &        &  O9-B5       & 1-3\,d      &        & NR\\
$\beta$ Cep & BCEP   &  O8-B6\,I-V & 0.1-0.6    & 0.01-0.3 & R + NR\\
            & BCEPS  &  B2-B3\,IV-V & 0.02-0.04  & 0.015-0.025 & R + NR \\  
SPB         & SPB    &  B2-B9\,V    & 0.4-5\,d    & $<$0.5 & high radial order, \\
            &        &              &            &         & low degree g-modes\\
Be          & BE, LERI &Be          & 0.3-3\,d   &         & NR (or rotational?)\\
LBV         & LBV     & hot SG      & 30-50\,d   &         & NR?\\
$\alpha$ Cyg & ACYG   & Bep-Aep\,Ia  & 1-50\,d   & $\sim$0.1 & NR, multiperiodic\\
BX Cir      &         & B           & ~0.1\,d    & $\sim$0.1 & H-deficient \\	   
PV Tel      & PVTELI  & B-A\,Ip     & 5-30\,d    & $\sim$0.1 & He SG, R strange mode\\
	    & PVTELII & O-B\,I      & 0.5-5\,d   &         & H-def. SG, NR g-mode \\     
	    & PVTELIII& F-G\,I      & 20-100 d	 &         & H-def. SG, R?\\
Fast Rotating & FaRPB & B           & $<$55\,d   & 0.001-0.004 & fast rotators\\
Puls. B Stars &&&&&\\[0.2ex]
\hline
Blue Large  & BLAP    & O-B         & 20-40 min  & 0.2-0.4 &\\
Ampl. Puls.  &&&&&\\[0.2ex]
\hline
Binary      &         &             &            &         &\\
Evolution   & BEP     &             &         &      & RR Lyr `impostors'\\
Pulsators   &         &             &         &      &  \\[0.3ex]
\hline
Heartbeat   &         &             &            &         &binary stars\\
Variables   &         &             &            &         &on eccentric orbit\\
\hline\hline
\end{tabular} 
\end{center}  
\end{table}	     

Oscillations are ubiquitous in stars. Hot and cool stars, luminous and
low luminosity stars can also pulsate, as is seen in the Hertzsprung-Russell 
(H-R) diagram showing the location of different types of pulsating variable 
stars (Fig.~1). Even our Sun is a pulsating variable star in which millions 
of non-radial oscillation modes have been excited.

Table~1 is an up-to-date overview of different types of pulsating 
variables whose oscillations are excited by various mechanisms.
In the last years, four new types of pulsating variables -- listed in the
end of Table~1 -- were discovered:\\
\noindent - fast rotating pulsating variables,\\
\noindent - blue large amplitude pulsators,\\
\noindent - binary evolution pulsators,\\
\noindent - heartbeat variables.

The first representatives of fast rotating pulsating stars were discovered
by Degroote et~al. (2009) and Mowlavi et~al. (2013). Stars belonging to
this new variability type also obey a period-luminosity ($P$-$L$) 
relationship but its cause differs from that of the ($P$-$L$) relationship
for classical pulsating variables (Mowlavi et al., 2016).

Blue large amplitude pulsators vary with short period like $\delta$~Scuti
stars but with larger amplitude and their effective temperature is
about 30\,000~K (Pietru\-kowicz et~al., 2017). Their luminosity corresponds 
to stars fainter than those of main sequence stars of similar temperature.
Such objects can be formed after merging of two low mass stars.

Binary evolution pulsators resemble RR~Lyrae type variables 
phenomenologically (this is why their nickname is RR~Lyrae impostors) but
their mass is lower (about half) than that of the RR~Lyraes on the
horizontal branch of the H-R diagram. Such stars can occur in the 
instability strip of classical pulsators as a result of mass transfer in
a binary system (Pietrzy\'nski et~al., 2012). Binary evolution pulsators 
can mimic Cepheid type behaviour, too. Based on a population synthesis 
calculation, Karczmarek et~al. (2017) concluded that about 0.8\% of 
seemingly RR~Lyrae type variables are in fact binary evolution pulsators,
while the contamination of Cepheids by such impostors is higher, about 5\%.

The heartbeat variables are composed of two stars on an eccentric orbit 
and tidal interaction excites low amplitude pulsation in at least one of
their components. The first such binary star was found in the Kepler field 
(Welsh et~al., 2011), and soon after this discovery, a number of such systems 
have been identified (Thompson et~al., 2012).

Pulsating variable stars are important objects for astrophysics, cosmology,
and studies of galactic structure, as well. From the point of view of 
astrophysics, stellar oscillations provide us with information on internal 
structure of the stars and their evolutionary phase. The cosmological use 
of pulsating variable stars is based on the fact that several types of 
luminous pulsators obey a specific $P$-$L$ relationship which is indispensable 
in extragalactic distance determination. The galactic structure can be traced 
by studying spatial distribution of various types of pulsating stars of 
different ages/populations within a galactic system. 

Importance of pulsating stars is supported by the frequency of occurrence
of such variables, as well. About 40\% of the designated variable stars 
in the General Catalogue of Variable Stars (GCVS, Samus et~al., 2017) 
are pulsating variables. Because of the strict criteria, GCVS designation 
has been assigned to only less than 60\,000 variable stars, meanwhile the 
number of known variables is over half a million. The Variable Star Index 
(VSX) maintained at the AAVSO web-site contains 543\,564 stars in October 
2018, while data on 550737 variable stars have been published 
in the Gaia Data Release~2 in April 2018, including more than 100\,000 
RR~Lyrae type variables and 10\,000 Cepheids (Gaia Collaboration, Brown 
et~al., 2018). Keeping in mind the fact that about 10\% of the targets of 
the Hipparcos astrometric satellite were found to be variable in brightness, 
the expected number of variable stars exceeds a hundred million among the
more than one billion targets of the ESA's Gaia project. Naturally, the 
number of pulsating stars is also extremely large among the Gaia sample of 
stars brighter than 20\fm7 magnitude.

When observing pulsating stars, temporal coverage (duration of the time 
series) is a critical aspect for studying multiperiodicity, changes in 
frequency content, modal amplitudes, etc. Therefore, time consuming 
photometry of pulsating variables is a realm of small telescopes.
If the astroclimate of the observing site does not allow high-precision
photometry, observations of variable stars with large photometric 
amplitude are recommended.

A tutorial on the basic notions related to stellar pulsation is
available in the pdf version of the author's presentation delivered
during the conference (Szabados, 2018).

\section{Importance of binarity among pulsating variables}

It is a remarkable fact that binarity is important in at least 
three of the four recently defined types of pulsators mentioned above. 
In addition, pulsating stars in eclipsing binary systems are invaluable 
sources of information because such pairs of stars offer a unique
possibility for an accurate determination of the mass, radius and 
luminosity of the components and test predictions of the pulsation theory
(see e.g. Pilecki, 2018). Moreover, bright companions can falsify the 
calibration of the $P$-$L$ relationship without correcting for their 
additional light. Therefore an important task is to reveal binary 
systems among pulsating stars used for standard candles.

A useful hint for binarity is the appearance of the light travel time
effect in the $O-C$ diagram due to the orbital motion in a binary
system. However, seemingly periodic variations in the pulsation period
are insufficient for declaring that the given pulsator is a member in a
binary system if the light-time effect solution results in unrealistic
stellar parameters or orbital elements as testified by the case of the
RR~Lyrae variable Z~CVn (Skarka et~al., 2018).

A close companion can even trigger stellar oscillations in the other star
of the binary system. This happens in the case of heartbeat variables.
Another kind of externally triggered pulsation was observed in the
symbiotic nova RR~Tel preceding its eruption in 1948 (Robinson, 1975).

Now it is clear that long-period variations in the mean brightness of 
RV~Tauri stars (RVB subtype) are also caused by the binarity of these 
pulsators (Kiss \& B\'odi, 2017, and references therein).

\section{Rapid evolutionary episodes in pulsating variables}

In this section, recently observed interesting examples of rapid 
evolutionary episodes are mentioned continuing the list published 
in Szabados (2014). 

The Type~II Cepheid, V725~Sgr, experienced a sudden transition  
to red semi-regular variable: its pulsation period increased
from about 10 days to 90 days within a century (Percy et~al., 2006).

The OGLE photometric survey is a treasure-house for finding
peculiar behaviour among various pulsators. For example,
Soszy\'nski et~al. (2014) revealed three RR~Lyrae type variables
which experienced mode switching from double-mode pulsation to
simple fundamental mode oscillation or vice versa. Their sample
consisted of about 38\,000 RR~Lyrae stars in the Galactic Bulge,
so such rapid transitions are not extremely rare.
Similarly, the OGLE project resulted in the discovery of a Cepheid 
variable, OGLE-SMC-CEP-3043, that stopped its pulsations within 15 
years of observation (Soszy\'nski et~al., 2015b).

Delta Scuti stars perform both radial and non-radial pulsations and
many modes can be excited simultaneously. It is noteworthy that 
temporal variations occur in both the observed frequencies and the
modal amplitudes on a time scale of months to years. A thoroughly
studied example is the case of AI~CVn (Breger et~al., 2017; Breger 
\& Lenz, 2018).

\section{New results on classical Cepheids}

Cepheids have been considered as extremely stable pulsators
for long. Studies based on very accurate photometric data 
collected by space telescopes, however, revealed that this
paradigm has to be revised. Cycle-to-cycle changes occur
in the pulsation period and light curve shape of V1154~Cygni 
(Derekas et~al., 2012). In spite of this slight period flickering, 
the average pulsation period has remained stable on the time 
scale of decades. From the analysis of Kepler data covering 
four years Derekas et~al. (2017) pointed out that the light 
curve of V1154~Cyg is modulated. This effect resembles the 
Blazhko effect commonly appearing among RR~Lyrae stars. In
the case of V1154~Cyg, the cycle length of the modulation is
about 159 days.

Blazhko effect was revealed in other classical Cepheids, too.
Berdnikov et~al. (2017) found strong Blazhko modulation in the
light curve of ASAS 160125-51503 with a cycle length of 1242
days. The pulsation of the best known Blazhko Cepheid, V473~Lyrae
(Moln\'ar et al., 2013) is subjected to period doubling, as 
revealed from uninterrupted photometry obtained with the MOST 
space telescope (Moln\'ar et~al., 2017). Such period doubling 
was recently observed in Blazhko RR~Lyrae stars, too.

A part of fundamental-mode Cepheids in both Magellanic Clouds 
have periodically modulated light curve (Smolec, 2017). Though
the amplitude of the modulation is tiny, the phenomenon is
intriguing.  

In addition to the slightly unstable light curve, the radial 
velocity phase curve of classical Cepheids is also 
non-repetitive because variations in the atmospheric velocity 
gradient result in radial velocity changes (Anderson, 2016). 

Most of Cepheids are single-mode pulsators, but simultaneously 
excited double-mode oscillation also exists among this type of
variable stars and even Cepheids pulsating simultaneously in 
three radial modes have been identified (Moskalik, 2013). 
Long-term modulation of the light curve has been observed among
the double-mode Cepheids of the Magellanic Clouds (Moskalik \&
Ko\l{}acz\-kowski, 2009). A characteristic feature of this 
modulation is that the amplitudes of modes involved vary in 
anticorrelated manner.

The period ratio of the firts overtone and the fundamental mode 
pulsation is sensitive to the atmospheric iron abundance of
double-mode Cepheids (Szil\'adi et~al., 2007). Recently Szil\'adi
et~al. (2018) published formulae allowing calculation of iron
abundance of double-mode Cepheids from the Fourier decomposition
of the light curve. 

The OGLE data show that additional periodic variability exists
among first overtone Cepheids in the Magellanic Clouds with a
photometric amplitude of 2-5 millimagnitude. The period is in the
range of 0.60-0.65 of the value of the main pulsation period 
(Smolec \& \.Sniegowska, 2016). In the Petersen diagram, three
well separated sequences are outlined corresponding to stars
that pulsate with this additional periodicity (see left panel of
Fig.~2). From the period ratio, the slightly excited oscillation
cannot be any radial mode. Pulsation with similar period ratio has
been observed in first overtone RR~Lyrae variables. According to
Dziembowski's (2016) theoretical calculations, the three new
sequences in the Petersen diagram correspond to stars pulsating
in non-radial modes of angular degrees 7, 8, and 9.

Anomalous Cepheids probably are old coalesced binaries
crossing the instability strip (Soszy\'nski et~al., 2015a).
Thus they are examples for binary evolution pulsators.
Because anomalous Cepheids have separate $P$-$L$ relationships,
their individual representatives can be readily identified in
the Magellanic Clouds based on the apparent mean brightness, 
unlike their Galactic counterparts.

\section{New results on Type~II Cepheids}

Type~II Cepheids are low-mass, usually metal-poor pulsators
oscillating in the radial fundamental mode. Their pulsation
is less stable compared with the oscillation of classical
Cepheids. Noticeable light-curve changes occur from one cycle
to the other, and the pulsation period can vary rapidly. In
addition to these long known interesting facts, other
peculiarities have been revealed quite recently (Smolec et~al. 
2018). The following dynamical phenomena have been identified
in Type~II Cepheids of the OGLE collections:\\
\noindent - double-mode pulsation in the short period (BL~Her
subgroup);\\
\noindent - period doubling behaviour in the BL~Her type
variables;\\
\noindent - quasi-periodic light-curve modulation in all three
(BL~Her, W~Vir, and RV~Tau) subgroups of Type~II Cepheids.\\
\noindent - Moreover, period doubling was revealed in Galactic 
W~Vir stars (Plachy, 2018).

DF~Cyg turned out to be the first RVb type variable star showing
low-dimensional chaos in its pulsation (Plachy et~al., 2018). 
This is only the third known case of chaotically pulsating
RV~Tauri stars, the other two variables belong to the RVa 
subgroup.

\section{New results on RR Lyrae type variables}

In addition to their periodic pulsation, the old RR~Lyr type 
variables situated on the horizontal branch of the H-R diagram
also show various other effects worthy of studying in detail. 
At the turn of the millennium, RR~Lyrae variables were thought
to be monoperiodic radial pulsators a part of which shows a
modulated light curve. Now we know that RR~Lyrae stars are
multiperiodic with radial and nonradial modes excited.
The long known though still mysterious phenomenon is the Blazhko 
effect, a slow, cyclic (not periodic) modulation of the light 
curve amplitude and phase appearing simultaneously is present in
about 50\% of these variable stars. It is typical that more than
one modulation cycle is present in a given star. A gallery of
Blazhko modulation is published by Benk\H{o} et~al. (2014).
It is promising that new models involving interactions between 
radial and nonradial modes of oscillation as well as coupling 
between the fundamental mode, first overtone and a high-order 
(9th) radial mode can lead us to the correct explanation of the 
Blazhko effect (Buchler \& Koll\'ath, 2011). Several months ago 
Zolt\'an Koll\'ath succeeded in calculating a model light curve 
resembling Blazhko modulated brightness variations.

\begin{figure}[!]
\centerline{\includegraphics[width=12.0cm,clip=]{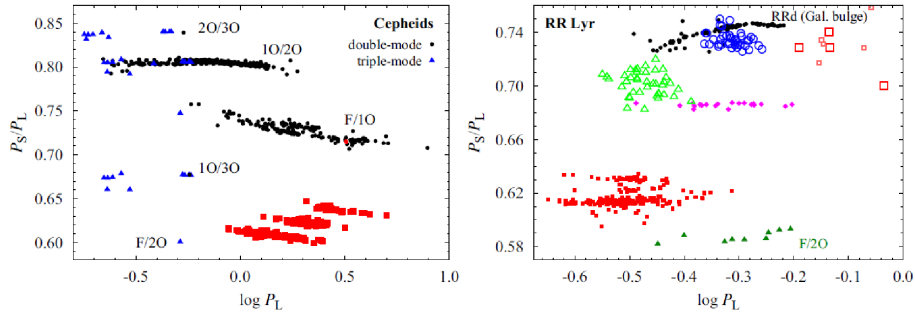}}
\caption{Petersen diagram for Cepheids (left) and RR~Lyrae stars (right)
 (Smolec et~al., 2017).}
\label{fig2}
\end{figure}

Double-mode pulsation and nonradial modes are also present in some
RR~Lyr type variables (Moskalik, 2013). Among the double-mode
RR~Lyrae variables in the Magellanic Clouds, Soszy\'nski et~al. (2016) 
identified anomalous RRd pulsators which are characterised by
unusual period ratios and modal amplitudes.

Based on Kepler and CoRoT data new dynamical phenomena have been 
discovered: period doubling and extra modes in all RR~Lyraes pulsating
in the first radial overtone, in all double-mode RR~Lyraes and all
Blazhko modulated fundamental pulsators (Netzel et~al., 2015).
These additional modes are non-stationary and their frequency
ratios concentrate in narrow intervals (see Fig.~2., right panel). 
The existence of extra nonradial modes is similar to the phenomenon
discussed in Section~4.

Another important aspect of the current RR~Lyrae research is the
quest for finding binaries among these old pulsators. There is only
one undoubted case: TU UMa (Li\v{s}ka et~al., 2016), unlike Cepheids
in our Galaxy where the occurrence of binaries exceeds 60\%.
The list of suspected binaries among RR~Lyrae stars can be found in
the online data base maintained by Li\v{s}ka \& Skarka.

\section{Telescopes, targets, tasks}

Observation of pulsating variable stars needs patience and huge
amount of time. Therefore small (up to 1.5\,m diameter) telescope 
are used for carrying out photometric observations of such variables.
Data bases of various photometric sky surveys also offer an opportunity
to study individual pulsating variable stars or a group of them.

\begin{table}[!] 
\begin{center}  
{\footnotesize
\caption{Space telescopes used for or dedicated to photometry} 
\label{spacephotom}  
\begin{tabular}{l@{\hskip0mm}c@{\hskip0mm}c@{\hskip1mm}l} 
\hline\hline
Mission (duration) &  Aperture & Band & Remarks\\
                   &   (cm)    & (nm) &\\
\hline  
IUE     (1978-1996) &  45  & wide & Fine Error Sensor, no calibration\\
 & & &0.03; var.; var.; targets for UV spectra \\
Hipparcos (1989-1993) &  29  & 400-800 ($Hp$)  & Tycho $B_T$: 350-500 nm; 
$V_T$: 460-600 nm\\
 & &  &0.02; 30-380; 3 years; 118\,000 (Tycho: $2.5\times 10^6$) \\
HST  (1990- )     & 240  & 460-700   & Fine Guidance Sensor\\
 & &  &0.001; var.; var.; millions \\
WIRE (1999-2011) &   5.2& $V+R_{\rm J}$& star tracker\\
 & &  & 0.005; $> 10^3$; 21\,d; 200 bright stars\\
INTEGRAL (2002- ) &   5  & Johnson $V$ & Optical Monitoring Camera \\
 & &  &0.04; var.; var.; $> 10^3$ \\
Coriolis (2003-2011) &   1.3& wide & on board SMEI satellite\\
 & & &0.02; $> 10^3$; years; all naked-eye stars \\
MOST (2003-2014) & 15 & 380-800  & limited field (CVZ)\\
 & &  &0.0001; $> 10^3$; up to 60\,d; 5000 bright stars \\
CoRoT (2006-2012) &  27  & 350-1000  & very limited field\\
 & &  &0.0001; $> 10^5$ ; 0.5 year; 120\,000 \\
Kepler (2009-2013) &  95  & 400-900   & very limited field\\
 & &  &0.00003; $> 10^5$ ; 4 years; 150\,000 \\
BRITE (2013/2014- ) &  3  & 550-700  & 5 satellites: 3 in blue, 2 in red band\\
 & & or 390-460 & 0.001; $> 10^3$; 30-180\,d; $\sim$1000 bright stars\\
Gaia (2013- )  &  68  & 330-1050 ($G$) & $G_{BP}$: 330-680 nm; $G_{RP}$: 640-1050 nm\\
 & & &0.001-0.020; $\sim$70; $>$5 years; 1.6 billion \\
K2 (2014-2018) &  95  & 400-900  & small fields along the Ecliptic\\
 & & & 0.0005; $10^5$ ; 80\,d; 350\,000\\
TESS (2018- )  &  4$\times$10.5  & 600-1000 & large field of view\\
 & &  & 0.0005; $> 10^4$ ; 27\,d; 200\,000 \\
CHEOPS (2019- ) &  30  & 400-1100 & to be launched\\
 & &  & 0.00002; n.a.; n.a.; bright exoplanet host stars \\
\hline\hline
\end{tabular} 
}
\end{center}  
\end{table}

In addition to ground-based equipments, space telescopes dedicated 
to or used for stellar photometry (see Table~2) also produce large 
amount of time series photometric data. Typical accuracy (in magnitudes), typical number of observations per target, length of the data series, 
and the approximate number of the observed stars are also listed in the 
second line of the remarks column of Table~2, for each space project. The 
advantage of space photometric data is twofold: their accuracy is much 
better than that of ground-based photometry, and the time series is
uninterrupted or there are only short gaps. This latter feature is 
important for identifying the oscillation frequencies present in the 
observed star.

In view of the tremendous number of recently discovered new variable 
stars and the ongoing survey of the ESA's astrometric space probe, Gaia,
chance of revealing new variables during our own observations is extremely
low. Observers with small telescopes cannot compete with the discovery
efficiency of the LSST, either (LSST Science Collab., 2012).
However, projects aimed at studying carefully selected individual pulsating
variables can be very productive. A not exhaustive list of the features 
to be investigated from the observational data is as follows:\\
\noindent - the value of the pulsation period can be updated using the 
$O-C$ method, if prior photometric data are available\\
\noindent - from detailed photometric study of individual variables 
one can point out additional periodicities, perform a mode 
identification, discover slightly excited non-radial (or radial) modes\\
\noindent - pulsating variables in binary systems can be especially useful 
targets because of chance of revealing interactions of binarity and pulsation
phenomena\\
\noindent - for Cepheids and RR~Lyrae stars, the atmospheric
metallicity can be determined from the shape of the light curve
via Fourier decomposition (Klagyivik et~al., 2013).

If photometric data are insufficient for a reliable analysis, an in-depth 
spectroscopic study with a larger telescope can be instrumental. In such
cases a cooperation between several telescopes/observatories is beneficial.

Many other interesting facts on pulsating variable stars are discussed 
in detail in the recent handbooks written by Aerts et~al. (2010), 
Balona (2010), Catelan \& Smith (2015), and Percy (2007), as well as in 
the conference proceedings edited by Su\'arez et~al. (2013).

\acknowledgements
The organizers of the conference are thanked for dedicating an invited
review to the topic of pulsating variables. This work has been supported 
by the Lend\"ulet Program of the Hungarian Academy of Sciences, project 
No. LP2018-7/2018 and the Hungarian NKFIH projects K-115\,709 and K-129\,249.

\bibliography{demo_caosp306}

%

\end{document}